\documentclass[nofootinbib, amsmath,amssymb, aps, twocolumn]{revtex4}

\usepackage{hyperref}

\usepackage{graphicx}
\usepackage{dcolumn}
\usepackage{bm}

\usepackage{booktabs}
\usepackage{topcapt}

\usepackage{amsmath}
\usepackage{amssymb} 
\usepackage[bbgreekl]{mathbbol}

\usepackage{xcolor}

\usepackage{slashed}

 \usepackage{ulem}
 \usepackage{cancel}

\begin{document}
\title{Using the Landau gauge gluon propagator to set the lattice physical scales \\ and understanding the finite size effects}
\author{Orlando Oliveira}
\email{orlando@uc.pt}
\author{Paulo J Silva}
\email{psilva@uc.pt}

\affiliation{CFisUC, Departament of Physics, University of Coimbra, 3004-516 Coimbra, Portugal}

\begin{abstract}
A crucial step in extracting physical predictions from lattice QCD simulations is the scale setting, 
i.e. the determination of the lattice spacing ($a$) in physical units. 
Herein, the relative scale setting for different $\beta$'s is discussed, using the Landau gauge gluon propagator computed with large statistical ensembles.
After setting the relative scales, finite size effects are observed in the ultraviolet regime and handled in an effective description,  inspired in perturbation theory. 
The new devised procedure is efficient in handling the finite size effects, linking the lattice simulations with continuum perturbation theory for the high momenta 
regime. Furthermore, the procedure can be extended to handle other Green functions computed within lattice QCD simulations.
\end{abstract}

\maketitle
\tableofcontents

\section{Introduction}

Lattice Quantum Chromodynamics (LQCD) simulations offer a first principles non-perturbative approach to solve QCD in Euclidean discretized spacetime using
a finite lattice. The simulations are performed with dimensionless quantities and, to compare observables with experimental values, a definition of a physical scale
is required. Once the physical scale is chosen, all other  quantities are measured relative to it. 
The scale setting can be translated into a definition 
of the lattice spacing $a$ or, equivalently, the lattice bare coupling constant.

In the early days of LQCD, the scale setting used the string tension or the $\rho$-meson mass as reference mass scale. 
The advent of improved actions, that reduced the discretizations errors, and the increase of the ensembles of configurations on a typical
simulation led to the introduction of other ways to set the scale such as the Sommer scale  \cite{Sommer1994}, that relies on the heavy quark potential, or
pseudoscalar decay constants and hadron masses. Modern estimations set the lattice spacing $a$ from Wilson flow scales  
\cite{Luscher2010}, such as $t_0$ or $w_0$, that allow for high precision determinations and reduced systematics.
Hadronic observables, like  $m_{\Omega}$ of $f_{\pi}$ are also widely used, see e.g. \cite{FLAG2021} and references therein.
Another possible choice is to combine multiple observables in global fits to define the lattice spacing $a$. 

There is no unique way to choose the lattice spacing and different methods result in different $a$ values, with differences of the order of few percent level.
The challenge for the scale setting  is to reconcile the different approaches, and to reduce the systematic uncertainties to arrive at high precision
QCD predictions.

For LQCD, the lattice spacing or, equivalently, the bare gauge coupling, is sensitive to the action discretization scheme and to the finite volume used in the simulation.
The renormalization of the theory removes the dependence on the cutoff, but  finite size effects and the breaking of the rotational invariance require
further considerations. Physical results require performing the continuum limit, that, in practice, can be difficult to realize.

The non-perturbative computation of the  two-point functions of the QCD fundamental fields, i.e. the gluon, the ghost and the quark propagators, using LQCD Monte Carlo
simulations are typical examples of two-point Green functions that have been studied. For the conversion into physical units, they all require the use of a scale that
can be defined from different quantities. The differences in the choice of the scale setting are within the order of a few percent.
For example, in \cite{Duarte:2016iko} the authors used the string tension to define the physical scale for the various $\beta$ values considered.
The observed differences led \cite{Boucaud:2017ksi,Duarte:2017wte,Boucaud:2018xup}  to discuss the scale setting for quenched Yang-Mills SU(3) theory.
In the last work the authors studied the gluon and ghost propagators, performed an analysis of the physical scaling violation of these propagators in the Landau
gauge and an extrapolation to the continuum limit. More, their work established a relative normalization of the scale setting, i.e. of the lattice spacing associated with
various $\beta$'s.

Herein, we revisit the problem of the scale setting for pure Yang-Mills SU(3) theory relying solely on the Landau gauge gluon propagator.
In this way, the computation
of the ghost propagator, that requires the inversion of a large sparse matrix, that is time consuming, is avoided. 
Furthermore, modern lattice QCD codes typically include a module to compute the Landau gauge gluon propagator.
The computation discussed relies on large 
ensembles of gauge configurations, of the order of 10K per ensemble, and on a minimization of the difference between propagators computed with different $\beta$, over
a finite range of momenta.

As discussed, the use of large ensembles results in smooth propagators, where the statistical fluctuations play essentially no role. A suitable reduction of
the differences between propagators within a given momentum range can be done and, with a proper choice of the relative scale, these differences can be eliminated
for the range of momenta considered. 
It turns out, that this relative scale setting brings an overall good agreement between the data generated with different $\beta$'s  --- however, small differences in the 
  UV regime persist. These differences at high momentum are due to the finite size effects in the numerical simulations.
 Such scale violations in the UV regime are also seen in previous studies, as can be observed in the data reported in e.g. \cite{Boucaud:2018xup}. 
One should alert the reader that the description of the propagator in the current work and in this later work do not rely  in exactly the same definitions. 
Probably the most important difference is that here the propagators are described in terms of the improved momentum, while in \cite{Boucaud:2018xup} only
the naive lattice momentum was considered; see below for definitions of the two momenta and comments on the choice of momenta.
The two definitions of momentum differ by $\mathcal{O}(a^2)$ corrections and, therefore, the corresponding lattice corrections are not expected to be the same.

The observed differences in the high momentum regime rise the question of the proper handling of the finite size and volume effects. 
By setting the relative scales by a minimization of the difference between propagators, the data reduces to a unique and well defined smooth curve up 
to momenta $\sim5$ GeV. The differences for higher momenta seem to be associated with finite size effects. 
As discussed in  \cite{Oliveira:2012eh}, the interplay of the finite lattice and the finite volume effects is far from being trivial.
Inspired in perturbation theory, we try to understand the scaling violations observed in the lattice data.
Our analysis explains, at the qualitative level and, to some  extent, at the quantitative level, the observed UV deviations between the different setups.
 The remaining observed small differences in the UV regime can be understood as an upper bound on the theoretical error associated with the calculation.

In the investigation of the scaling violations a comparison of lattice data in the UV regime with continuum renormalization improved perturbation theory 
is done. For the comparison with the continuum perturbation theory prediction both $\Lambda_{QCD}$ and the strong coupling constant $\alpha_s ( \mu)$, 
at the renormalization scale considered, have to be used. A possible choice is to take these values from the literature. 
Another way, is to measure $\Lambda_{QCD}$ and $\alpha_s ( \mu)$ from the lattice data. However, in  practice, to be able to measure these quantities from the
simulations, an understanding of the finite size effects is required. 
Our study finds that the functional forms inspired in continuum perturbation theory fail to reproduce the high momentum lattice data. 
On the other hand, we find that the UV regime of the lattice data, where the scale violation occurs,
is well reproduced by combining the results of the one-loop stochastic lattice perturbation theory, extrapolated to the continuum limit, with an effective way of 
describing the finite size effects. The functional form describing effectively the finite size effects can be understood from 
perturbation theory. Moreover, the technique used to define  the functional form, and to handle the finite size effects, can be generalized to other Green functions 
as, for example, the quark propagator, where it is known that finite size effects have sizeable contributions to the lattice propagator \cite{Oliveira:2018lln}.

In summary, in this work a simple way of setting the relative scale associated with different $\beta$'s for lattice QCD simulations, that relies on the measurement of the Landau gauge
gluon propagator, is provided, together with a  way to handle the finite size effects in the UV regime. 
Before proceeding, we recall the reader, that different setups, with different definitions, need its proper handling of all quantities
and the procedure described has to be tuned for each of the definitions.

This work is organized as follows. In Sec. \ref{Sec:Def} we provide the basic definitions used throughout the current work.
The details of the simulations and ensembles generated are reported in Sec. \ref{Sec:Setup}.
The setting of the relative scaling is worked out in Sec. \ref{Sec:Relative}, while the discussion of the scaling violations relying in the ultraviolet regime and relying in
perturbation theory is reported in Sec. \ref{Sec:Art}. Finally, in Sec. \ref{Final} we summarize and conclude.

\section{Definitions \label{Sec:Def}}

The gluon propagator in the Landau gauge is accessed from the evaluation of the
two point correlation function
\begin{equation}
 \langle A^a_\mu (p^\prime) \, A^b_\nu (p) \rangle = V \, \delta (p^\prime + p) \, D^{ab}_{\mu\nu} (p)
  \ ,
\end{equation}
where $\langle \cdots \rangle$ stands for the vacuum expectation value that, in a lattice simulation,
is realised as an ensemble average. 
The Slavnov-Taylor identity associated with the gluon propagator requires that, in the Landau gauge,
$D^{ab}_{\mu\nu} (p)$ is orthogonal to the gluon momentum, i.e. $p_\mu D^{ab}_{\mu\nu} (p) = p_\nu D^{ab}_{\mu\nu} (p) =0$.
The tensorial structure of the propagator is then given by
\begin{equation}
 D^{ab}_{\mu\nu} (p) = \delta^{ab} \left( \delta_{\mu\nu} - \frac{p_\mu p_\nu}{p^2} \right) D(p^2) \ .
\end{equation}

In a lattice simulation, due to the breaking of rotational symmetry, the dependence of the propagator $D(p^2)$ is replaced by a more
elaborated function, whose dependence on the momentum becomes 
complex \cite{Leinweber:1998im,Leinweber:1998uu,Becirevic:1999uc,deSoto:2007ht,Vujinovic:2018nqc,Catumba:2021hcx}.
An extrapolation to the continuum limit should be performed to resolve these issues.
Furthermore, the fulfillment of the orthogonality condition both for the propagator, i.e.  $p_\mu D^{ab}_{\mu\nu} (p) = 0$,
and for the gluon field, i.e. $p_\mu A_\mu(p) = 0$ as demanded by the Landau gauge fixing condition,
depends on the choice of what is taken as momentum \cite{Leinweber:1998uu}. Herein, the gluon field is given by
\begin{equation}
A_\mu (x + a \hat{\mu}/2)  = \left. \frac{ U_\mu (x) - U^\dagger_\mu (x) }{2 \, a \, g_0} \right|_{\mbox{Traceless}} \ ,
\end{equation}
where $a$ is the lattice spacing, $g_0$ the bare lattice coupling constant, $\hat{\mu}$ the unit vector along the lattice direction $\mu$,
and $U_\mu$  stands for the gauge link.
Following the notation of \cite{Silva:2004bv}, where the reader can find further details on definitions and on the gauge fixing, 
the momentum space gluon field is
\begin{equation}
  A_\mu(q) = \sum_x ~e^{- i q ( x +  a \hat{\mu}/2 )} ~ A_\mu (x + a \hat{\mu}/2)  \ ,
\end{equation}
where the naive momentum reads
\begin{equation}
q_\mu = \frac{ 2 \, \pi}{a \, L_\mu } ~ n_\mu 
\qquad  , \qquad n_\mu = -L_\mu/2+1, \dots, L_\mu/2
\label{p:naive}
\end{equation}
with $L_\mu$ being the number of lattice points along direction $\mu$. It is common to consider the improved momentum
\begin{equation}
p_\mu = \frac{ 2}{a} \,\sin \left( \frac{\pi}{\, L_\mu } ~ n_\mu  \right)  = q_\mu + \mathcal{O}(a^2) \ .
\label{p:improved}
\end{equation}
From the numerical perspective \cite{Leinweber:1998uu}, the orthogonality conditions are better fulfilled by using the improved momentum instead of the
naive definition. Moreover, in lattice perturbation theory the tree level gluon propagator is a function of the improved momentum only.
The replacement of $D(q^2)$ by $D(p^2)$ improves both the orthogonality conditions and reduces the effects due to the breaking of the rotational group
$O(4)$ in the lattice data. However, in what concern the finite size effects they are clearly visible in $D(p^2)$ \cite{Leinweber:1998uu,Vujinovic:2018nqc,Catumba:2021hcx}.
A common practise to reduce/eliminate the finite size effects is to consider only special sets of momenta, the momenta with minimal effects coming from the 
breaking of the $O(4)$ symmetry, to represent the propagator $D(p^2)$. The choice of these momenta was discussed in \cite{Leinweber:1998uu} and will be referred here as
momentum cuts. All the data to be reported below consider only those momenta that comply with the cuts defined in this work. The exceptions will be clearly stated.

\section{Lattice setup \label{Sec:Setup}}

The lattice simulations  considered here look at the following setups
\begin{itemize}
\item[I.]
$32^4$ lattice for $\beta = 6.0$ and $48^4$ lattice for $\beta = 6.2$,

\item[II.]
$48^4$ lattice for $\beta = 6.0$ and $64^4$ lattice for $\beta = 6.2$.
\end{itemize}
In each of the two sets of configurations, the two lattices have different lattice spacing but approximately the same physical volume. The details of
each ensemble can be found in Tab. \ref{tab:setup}, together with two estimations of the lattice spacing from different techniques that will be used
to convert to physical units. The lattices in setup I have a physical volume $\sim (3.1$ fm$)^4$, while those in the setup II have a physical volume
of $\sim (4.5$ fm$)^4$.

\begin{table*}[t]
   \centering
   \begin{tabular}{l@{\hspace{0.75cm}}r@{\hspace{0.5cm}}r@{\hspace{0.75cm}} r@{\hspace{0.5cm}} r@{\hspace{0.5cm}} r  @{\hspace{1cm}}  r@{\hspace{0.5cm}} r@{\hspace{0.5cm}} r } 
      \hline
      \hline
      $\beta$    & $L^4$ & \#  Confs      &  $a$                       & $a^{-1}$ & $aL$ & $a$ fm                                  & $a^{-1}$ & $aL$  \\
                      &            &                     & fm   &      GeV   &     fm           & &  GeV                 &    fm\\
      \hline
      \hline
      6.0 & $32^4$ & 10000 & 0.1016   & 1.943 & 3.25    & 0.096 & 2.056 & 3.07\\
      6.0 & $48^4$ & 9035   &  0.1016  & 1.943 & 4.88    & 0.096 & 2.056 & 4.61 \\
      6.2 & $48^4$ & 10000 & 0.07261 & 2.718 & 3.49    & 0.070 & 2.819 & 3.36 \\
      6.2 & $64^4$ & 10000 & 0.07261 & 2.718 & 4.64    & 0.070 & 2.819 & 4.48 \\
      \hline
   \end{tabular}
   \caption{Details of the simulations, including the total number of gauge configurations in each ensemble. The first set of numbers for the lattice spacing
   is taken from \cite{Oliveira:2012eh},
   while the second set is taken from \cite{Pinto-Gomez:2024mrk}. In both cases possible errors on $a$ were ignored. The table listed in \cite{Oliveira:2012eh}
   refer only to the statistical errors on $a$ that are below 2.5\%.}
   \label{tab:setup}
\end{table*}

The simulations reported were performed using the Chroma \cite{Edwards:2004sx} and PFFT \cite{PFFT} 
libraries. The gauge fixing was performed with a Fourier accelerated steepest descent
method, and uses as stopping criterion the lattice averaged Landau gauge condition that should be below $10^{-15}$ in lattice units. 
In \cite{Silva:2004bv} the reader can look for definitions and details on the gauge fixing algorithm.
The statistical errors were computed with the bootstrap method with a 67.5\% confidence level.

The bare lattice data defined by the (cylindrical plus conical)  momentum cuts, for all lattices, is reported in Fig. \ref{fig:baredressing} in terms of dimensionless quantities.
The large statistical ensembles considered in the evaluation of propagators define smooth curves that, in terms of dimensionless quantities, overlap
within each $\beta$ value. The statistical errors are rather small, that makes the fitting to functional forms difficult, but enable the use of interpolation
to estimate intermediate momentum values not accessed directly in the simulation.
The comparison of the data associated with the two $\beta$ values considered requires the renormalization of the bare data and, therefore,
a choice of scale for each simulation. It is expected the renormalization to remove the dependence on the cut-off, here defined by the lattice spacing.

\section{Setting the relative scale \label{Sec:Relative}}

\begin{figure}[t]
   \centering
   \includegraphics[scale=0.32]{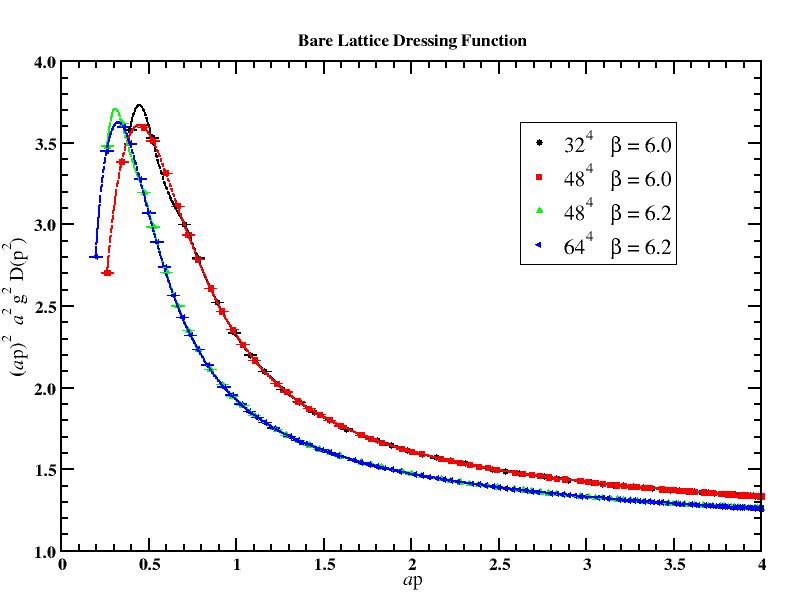} 
   \caption{The dimensionless lattice bare dressing function $p^2 D(p^2)$. The data reported here complies the cylindrical plus conical momentum cuts as defined in
                 \cite{Leinweber:1998uu}. The full lines are the curves that interpolate, with cubic splines, the lattice data points. See text for details.}
   \label{fig:baredressing}
\end{figure}

For the comparison of the data associated with the two $\beta$ values the lattice data points are interpolated using cubic splines. In all cases, the
zero momentum data is not taken into account in the interpolation. 
The large number of gauge configurations in each ensemble,  enables the use of interpolation to access the propagators at all intermediate momenta. 
In Fig. \ref{fig:baredressing}{\color{blue},} besides the bare data points for the dressing functions, it includes the curves that interpolate the lattice data.

\begin{figure}[t]
   \centering
   \includegraphics[scale=0.31]{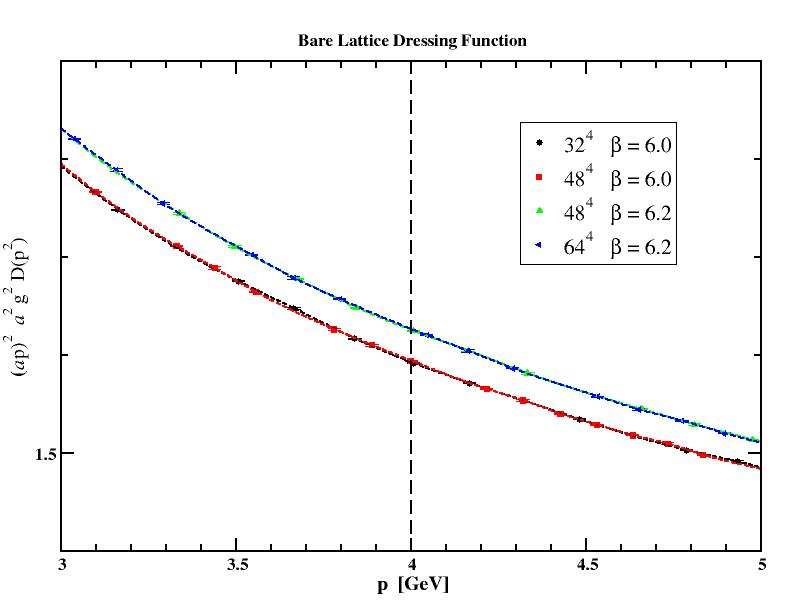} 
   \caption{The lattice bare gluon propagator in terms of the improved momenta in GeV. In the conversion into physical units the scale setting as in \cite{Oliveira:2012eh} was used. 
   The figure also shows the interpolation of the lattice data. The dashed vertical lines corresponds    to the momenta used to renormalize $D(p^2)$.}
   \label{fig:baredressing-pGeV}
\end{figure}

\begin{figure}[t]
   \centering
   \includegraphics[scale=0.31]{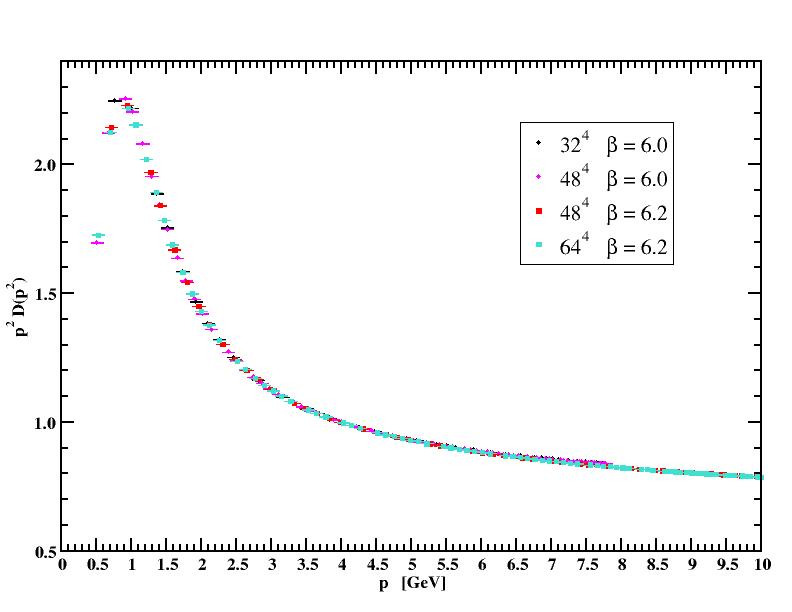} 
   \caption{The lattice  dressing function renormalized at $\mu = 4$ GeV for all the data sets. For the conversion to physical units we used
                 for the lattice spacing the numbers quoted in \cite{Oliveira:2012eh}; see Tab. \ref{tab:setup}.}
   \label{fig:R4GeVdressing}
\end{figure}

The outcome of the interpolation is  used to renormalize the lattice data in the MOM scheme, i.e. requiring $D(\mu^2) = 1 / \mu^2$. 
The renormalization scale considered herein is $\mu = 4$ GeV. The renormalized data using the scale setting as defined in \cite{Oliveira:2012eh}
is given in Figs \ref{fig:baredressing-pGeV} and \ref{fig:R4GeVdressing}. In both figures
the values for $a^{-1}$  used ignore any errors on the lattice spacing that can come either from
the finite statistics or systematics of the method used to defined the physical scales. For the values quoted in \cite{Oliveira:2012eh} the statistical
errors are of the order of 2.5\% or below. Similar curves can be build using the scaling setting considered in \cite{Pinto-Gomez:2024mrk}.

\begin{figure}[t]
   \centering
   \includegraphics[scale=0.31]{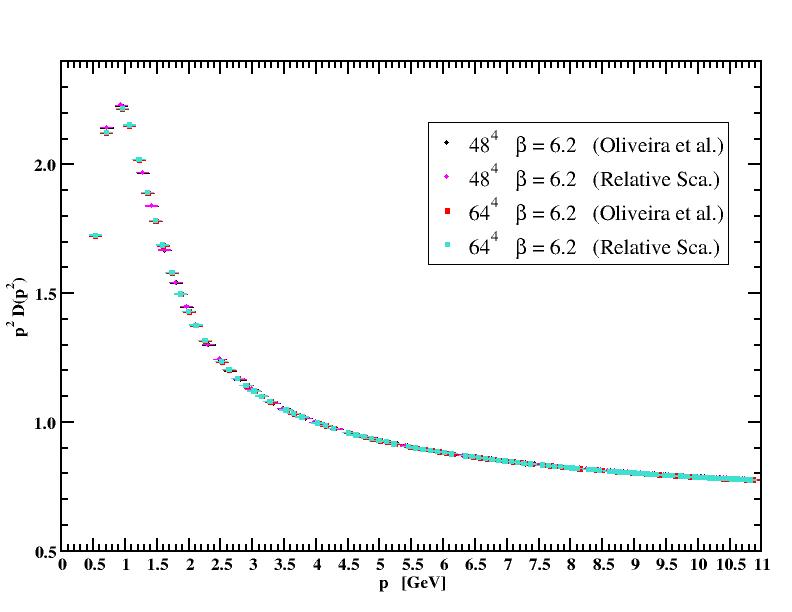} 
   \caption{The lattice  dressing function, renormalized at $\mu = 4$ GeV, for the $\beta = 6.2$ data sets, using the relative scale that sets
                  $1/a(\beta = 6.0) = 1.943$ GeV and $1/a(\beta = 6.2) = 2.7052$ GeV.}
   \label{fig:R4GeVdressing-beta6p2}
\end{figure}

\begin{figure*}[t]
   \centering
   \includegraphics[scale=0.31]{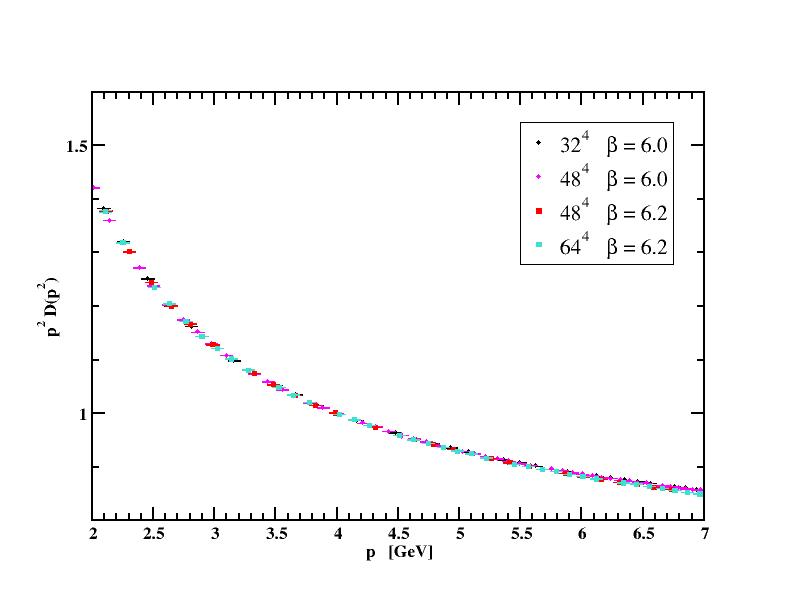} 
   \includegraphics[scale=0.31]{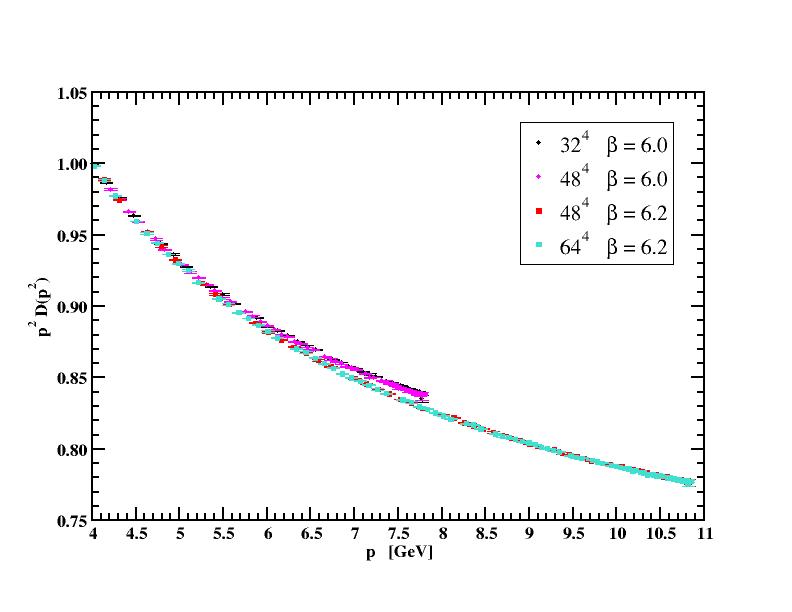} 
   \caption{The lattice  dressing function, renormalized at $\mu = 4$ GeV, for all data sets, using the relative normalization that sets
                  $1/a(\beta = 6.0) = 1.943$ GeV and $1/a(\beta = 6.2) = 2.7052$ GeV.}
   \label{fig:R4GeVdressing-UV}
\end{figure*}

In order to compare the lattice data from the two  lattice setups, and remove any remaining dependence on the cut-off, inspired on \cite{Boucaud:2018xup}
the relative scale is adjusted to improve the agreement between the two data sets for the same physical volume. This is performed minimizing
the difference
\begin{equation}
\delta = \int^{p_{max}}_{p_{min}} dp ~\Big( D^{(\beta = 6.0)}(p^2) - D^{(\beta = 6.2)}(p^2) \Big)^2
\end{equation}
using renormalized propagators. This was done independently for the two physical volumes described in Tab. \ref{tab:setup}. The motivation for introducing $p_{min}$ is to avoid 
bias coming from the interpolation due to the reduced number of points that survive the momentum cuts at low momenta. In our analysis we consider
$p_{min} = 1.5$ GeV. On the other hand, the rationale to have a maximum $p$ in the definition of $\delta$ is to avoid possible effects whose origin is
either the unresolved dependence on the cut-off or any other  effects that come from the breaking of rotational symmetry.
The minimization was performed considering $p_{max} = 7$ GeV and $p_{max} = 5$ GeV but no difference in the relative scaling between
the two $\beta$ values was observed. In the following we report on the results computed with the largest momentum range, i.e. with $p_{max} = 7$ GeV.

For setting the relative scale it was defined $a^{-1}(\beta = 6.2) = \lambda ~ a^{-1}(\beta = 6.0)$, then the bare lattice data was renormalized
as described previously and, only then, the integrated difference $\delta$ was computed. The $\lambda$ values that minimize $\delta$ are
$\lambda = 1.3928$ for smaller lattice volume and $\lambda = 1.3918$ for the larger lattice volume when one uses the numerical value of $a^{-1}$ for $\beta=6.0$ 
 as in 
\cite{Oliveira:2012eh}, while for the $a^{-1}(\beta=6.0)$ used in \cite{Pinto-Gomez:2024mrk} the corresponding values are
$\lambda =  1.3942$ and $\lambda = 1.3943$, respectively. These numbers should be compared with the actual scale ratios from 
\cite{Oliveira:2012eh} that is $\lambda = 1.399$, and $\lambda = 1.3714$ from \cite{Pinto-Gomez:2024mrk}. For both cases,
the relative scale for the two physical volumes are in excellent agreement, with average values of
$\lambda = 1.3923$, using the value of the lattice spacing from \cite{Oliveira:2012eh} for $\beta = 6.0$, and
$\lambda = 1.39425$, when using the value of the lattice spacing from \cite{Pinto-Gomez:2024mrk} for $\beta = 6.0$.
The corresponding $\beta = 6.2$ inverse lattice spacings are 2.7052 GeV, to be compared with 2.718 GeV (a difference of 0.5\%), for \cite{Oliveira:2012eh} 
and 2.8659 GeV, to be compared with 2.819 GeV (a difference of 1.6\%), for \cite{Pinto-Gomez:2024mrk}.

From now on we will take $1/a(\beta = 6.0) = 1.943$ GeV and $1/a(\beta = 6.2) = 2.7052$ GeV for the conversion into physical units. Note that the renormalized propagators that use this relative  scale setting are, at naked eye, undistinguishable from those in Fig. \ref{fig:R4GeVdressing}.
In fact, Fig. \ref{fig:R4GeVdressing-beta6p2}  compares the $\beta = 6.2$ lattice dressing functions using the relative scales based on 
\cite{Oliveira:2012eh}  and the ones computed here.

In Fig. \ref{fig:R4GeVdressing-UV} a zoom over the mid and highest momenta accessed in the simulations is given. There is clearly a dependence on
the cut-off that is not removed by the renormalization. This difference shows up only for $p \gtrsim 5.5$ GeV. A similar effect can
be observed in the  gluon propagator data published in \cite{Boucaud:2018xup}. This can only be an effect that is due to the breaking of rotational
invariance, i.e. a remnant of the lattice artefacts.

\section{Lattice Artefacts \label{Sec:Art}}

The understanding of the different behaviour, with $\beta$, observed at higher momenta for the renormalized propagator in Fig. \ref{fig:R4GeVdressing-UV} 
can be explained with the help of perturbation theory. At large momenta, one expects the lattice propagator $D_L$ to be described by 
\cite{Kawai:1980ja,DiRenzo:2009ni,DiRenzo:2010cs}
\begin{equation}
  D_{L}(p^2) = D_{C}(p^2)  \left( 1 + \sum_j c_j (a, V)  \left( \ln \frac{p^2}{\zeta^2} \right)^j \right)  ,
  \label{Eq:LatCorr}
\end{equation}  
where $D_C$ is the continuum propagator, the coefficients $c_j( a, V)$ are proportional to $\alpha_s^j (\mu)$,
with $\alpha_s (\mu)$ being the strong coupling constant at the renormalization scale $\mu$.
$\zeta$ is a mass scale that we will identify with the inverse of the lattice spacing. 

\begin{figure*}[t]
   \centering
   \includegraphics[scale=0.31]{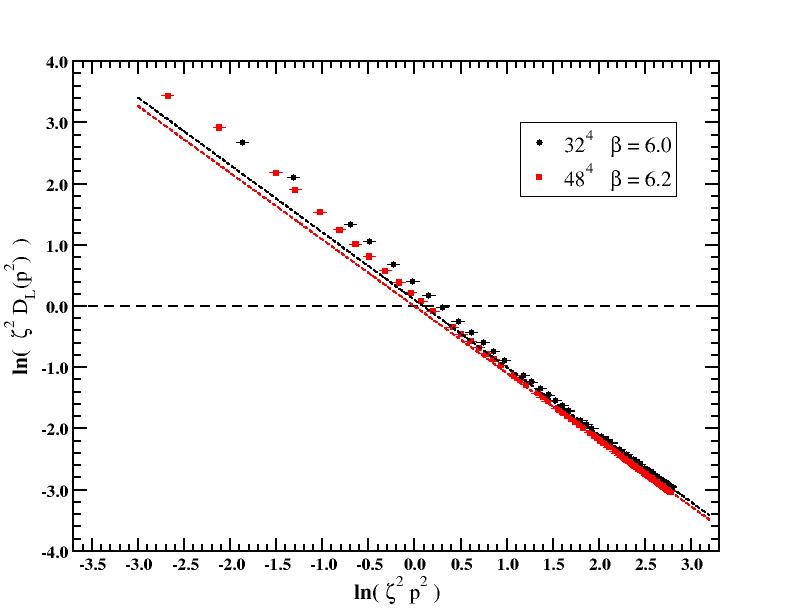} 
   \includegraphics[scale=0.31]{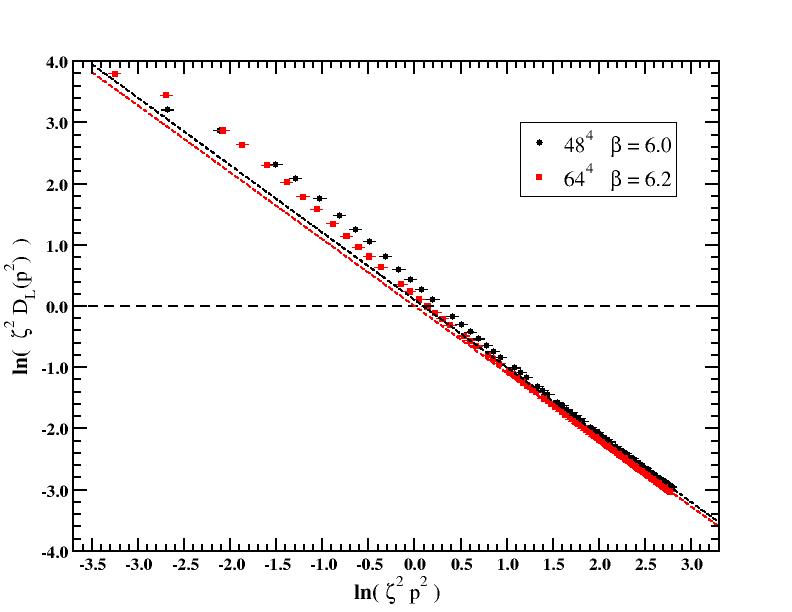} 
   \caption{Log-Log plot for the renormalized lattice data. The upper plot refers to the data computed using smaller physical volume, while the lower plot refers
                 to the largest volume. The dashed lines are the outcome of a linear fit.}
   \label{fig:R4GeVLogLog}
\end{figure*}

The use of the above expression to evaluate the lattice artefacts require the knowledge of $D_{C}(p^2)$. The asymptotic behaviour of the
renormalization group improved perturbation theory for  the gluon propagator at one-loop reads
\begin{equation}
D_C(p^2) = \frac{Z (\mu )}{p^2} \left( 1 + \omega \ln \frac{p^2}{\Lambda^2} \right)^{- \gamma_{glue}} \ ,
\label{Eq:ContProp}
\end{equation}
where $Z (\mu)$ is a constant that depends on the renormalization constant, $\omega = 33 \, \alpha_s (\mu) / 12 \, \pi$ for pure gauge 
SU(3) Yang-Mills theory, $\Lambda$ is the usual  QCD scale and $\gamma_{glue} = 13/22$ is minus the gluon anomalous dimension. 
The constant $\omega$ can be estimated using
the results of \cite{Aguilar:2010gm} that gives $\alpha_s(4 \mbox{GeV}) = 0.309$, with a $\Lambda = 0.425$ GeV, and results in $\omega = 0.2705$. 
These are the values to be used here in any numerical estimation.
For the precision achieved in our simulations, it is impossible to fit the lattice data to Eq. (\ref{Eq:ContProp}) with the expected value for
$\omega$ or close to it. However, this expression suggests that 
\begin{widetext}
\begin{eqnarray}
\ln \zeta^2  D_{L}(p^2) & = & 
    \ln Z (\mu ) 
    - \ln\frac{p^2}{\zeta^2}
    - \gamma_{glue} ~ \ln \left( 1 + \omega \, \ln \frac{p^2}{\Lambda^2} \right)
    + \ln \left( 1 + \sum_j c_j (a,V) ~ \left( \ln \frac{p^2}{\zeta^2} \right)^j \right) 
    \nonumber \\
        & = &
    \ln Z (\mu )     
    +   \Big( c_1 (a, V) - 1 - \gamma_{glue}  ~ \omega \Big)~ \ln \frac{p^2}{\zeta^2} 
    + \mathcal{O} ( \alpha_s^2 )
    \label{Eq:ExpAlpha}
\end{eqnarray}
\end{widetext}
and the curve $\ln \zeta^2  D_{L}(p^2)$ as a function of $\ln p^2/\zeta^2$ is expected to be linear or almost linear in the UV region.
Indeed, as seen in Fig. \ref{fig:R4GeVLogLog}, where the errors were computed assuming Gaussian error propagation and $\zeta = 1.943$ GeV, 
this function looks linear at naked eye. 
To quantify this statement, linear fits taking into account the statistical errors on the data, but ignoring correlations between the various data points,
were performed.
The fits with a $\chi^2/d.o.f.$ closer to the identity are represented by the dashed lines in Fig. \ref{fig:R4GeVLogLog} and their outcome is summarized in Tab. \ref{tab:LogLogFits}.

\begin{table*}[t]
   \centering
   \begin{tabular}{l@{\hspace{0.75cm}}r@{\hspace{0.75cm}} c@{\hspace{0.5cm}}  c @{\hspace{0.75cm}} c @{\hspace{0.75cm}} r @{\hspace{0.5cm}} l @{\hspace{0.75cm}} l} 
      \hline
      \hline
      $\beta$    & $L^4$ & \#  $d.o.f.$      &  $\chi^2/d.o.f.$   & Range & \multicolumn{1}{l}{$a$} & \multicolumn{1}{l}{$b$} & $c_1(a,V)$\\
      \hline
      \hline
      6.0           & $32^4$ & 13                  & 0.88                  & 2.3 - 3.0 & 0.0987(67) & -1.0991(25) & 0.0607(25) \\
      6.2           & $48^4$ & 25                 &  1.00                  & 2.5 - 3.0 & -0.0028(25) & -1.08989(93) & 0.06991(93)\\
      \hline
      6.0          & $48^4$ & 19                   & 1.15                  & 2.6 - 3.0 & 0.1012(66) & -1.1004(24)    & 0.0594(24) \\
      6.2          & $64^4$ & 36                   & 1.08                  & 2.5 - 3.0 & 0.0012(24) & -1.09155(90) & 0.06825(98)\\
      \hline
   \end{tabular}
   \caption{Linear fit to $a + b \, \ln (p^2/\zeta^2)$ whose value of $\chi^2/d.o.f.$ is close to unity. The estimation of $c_1$ uses $\omega = 0.2705$ and relies
   on the first order approximation in $\alpha_s$ to (\ref{Eq:ExpAlpha}). We recall that in this expression the perturbative one-loop renormalization group improved 
   approximation to the gluon propagator, that fails to reproduce the lattice data, is used. See main text for details.}
   \label{tab:LogLogFits}
\end{table*}

The fits estimations of the parameter $b$ are not compatible for the two  $\beta$ values. The data suggests that the $b$ associated with
the larger $\beta$ is smaller, as expected from naive considerations.  Note, however, that the estimation of $c_1 = 1 - \omega$, with $\omega = 0.2705$,
relies in an analytical approximation to the gluon propagator that is not able to reproduce the lattice data. This is possibly the reason why $c_1$ seems
to increase, instead of decreasing as approaching the continuum limit, i.e. as $\beta$ is increased.

\subsection{An effective description of the lattice artefacts}

\begin{table*}[t]
   \centering
   \begin{tabular}{l@{\hspace{0.25cm}}r@{\hspace{0.5cm}} c@{\hspace{0.5cm}}  c @{\hspace{0.5cm}} l @{\hspace{0.75cm}} l @{\hspace{0.5cm}} l @{\hspace{0.75cm}} l} 
      \hline
      \hline
      $\beta$    & $L^4$ & \#  $d.o.f.$      &  $\chi^2/d.o.f.$   & Range & \multicolumn{1}{l}{$Z$} & \multicolumn{1}{l}{$c_1$} & $c_2$\\
      \hline
      \hline
      6.0           & $32^4$ & 25                  & 0.99                  & 5.5 - ~\,8.0 & 0.5059(11)   & $( - 0.82 \pm 8.57 ) \times 10^{-5}$ & $6.97(97) \times 10^{-6}$ \\
      6.2           & $48^4$ & 44                 &  0.91                  & 6.8 - 11.0 & 0.45724(42) & $2.69(18) \times 10^{-4}$ & $(7.6 \pm 1.1) \times 10^{-7}$\\
      \hline
      6.0          & $48^4$ & 35                   & 1.08                  & 5.9 - ~\,8.0 & 0.5023(13)     & $(2.4 \pm 1.0) \times 10^{-4}$    & $(4.1 \pm 1.1) \times 10^{-6}$ \\
      6.2          & $64^4$ & 62                   & 1.03                  & 6.8 - 11.0 & 0.45707(35) & $2.76(16) \times 10^{-4}$   & $ 6.94 (98) \times 10^{-7}$ \\
      \hline
   \end{tabular}
   \caption{Parameter estimation fitting the lattice propagator data to Eq. (\ref{Eq:EffParLat}). $c_1$ has dimensions of $M^{-2}$, while
                 $c_2$ has dimensions of $M^{-4}$, where $M$ is a mass scale.}
   \label{tab:EffArt}
\end{table*}

\begin{table*}[t]
   \centering
   \begin{tabular}{l@{\hspace{0.25cm}}r@{\hspace{0.5cm}} c@{\hspace{0.5cm}}  c @{\hspace{0.5cm}} l @{\hspace{0.75cm}} l @{\hspace{0.5cm}} l @{\hspace{0.75cm}} l @{\hspace{0.75cm}} l} 
      \hline
      \hline
      $\beta$    & $L^4$ & \#  $d.o.f.$      &  $\chi^2/d.o.f.$   & Range & \multicolumn{1}{l}{$Z$} & \multicolumn{1}{l}{$c_1$} & $c_2$ & $c_3$\\
      \hline
      \hline
      6.0           & $32^4$ & 27     & 1.00     & 5.0 - ~\,8.0 & 0.5247(27)    & -0.00237(37) & $6.15(93) \times 10^{-5}$ & $ -3.95(72) \times 10^{-7}$\\
      6.2           & $48^4$ & 49     &  0.97    & 5.9 - 11.0   & 0.46561(80)  & -0.000332(66) & $8.48(94) \times 10^{-6}$ & $-3.06 (41)\times 10^{-8}$\\
      \hline
      6.0          & $48^4$ & 44     & 1.05      & 4.8 - ~\,8.0 & 0.5243(18)    & -0.00242(26)    & $6.39(66) \times 10^{-5}$  & $-4.22(53) \times 10^{-7}$\\
      6.2          & $64^4$ & 68     & 1.04      & 6.0 - 11.0    & 0.46570(76) & -0.00035(62)    & $ 8.76(87) \times 10^{-6}$ & $-3.20(37)\times 10^{-8}$ \\
      \hline
   \end{tabular}
   \caption{Parameter estimation fitting the lattice propagator data to Eq. (\ref{Eq:EffParLat}). $c_1$ has dimensions of $M^{-2}$, 
                 $c_2$ has dimensions of $M^{-4}$, and $c_3$ has dimensions of $M^{-6}$ where $M$ is a mass scale.}
   \label{tab:EffArtp4}
\end{table*}

As any lattice simulation, those discussed in this work access a limited range of momenta. In the $\beta = 6.0$ simulations the maximum momentum
achieved is about $8$ GeV, while for $\beta = 6.2$ the range of momenta goes up to $\sim 11$ GeV. 
To describe the high momenta region accessed in the simulations, and where the scale violations take place, we rely on
the results reported in \cite{DiRenzo:2010cs,Kawai:1980ja,Gracey:2003yr}, 
where the lattice Landau gluon propagator is evaluated with numerical stochastic perturbation theory on the lattice.
There, the continuum limit of the $\beta$ dependent gluon propagator is computed up to one-loop and reads
\begin{equation}
D(p^2) = Z \, \frac{ 2.29368 - 0.24697\,  \ln \left( a^2p^2 \right) }{p^2} \ .
\label{Eq:EffParLatCont}
\end{equation}
Following \cite{Ilgenfritz:2007qj} the lattice gluon propagator is described by
\begin{eqnarray}
D(p^2) & = & Z \, \frac{ 2.29368 - 0.24697\,  \ln \left( a^2p^2 \right) }{p^2} 
\nonumber \\
& & 
\quad
+ ~ c_1 ~ + ~ c_2 \, p^2 ~ + ~ c_3 \, p^4 ~ + ~  \dots\ ,
\label{Eq:EffParLat}
\end{eqnarray}
where the first term corresponds to the one-loop lattice stochastic perturbation  theory extrapolated to the continuum limit, 
and the coefficients $c_i$ parametrize the lattice artefacts. 

The motivation for using Eq.  (\ref{Eq:EffParLat}) to describe the lattice data, including lattice artefacts, can be understood as follows. In the continuum
formulation, the tree level bosonic propagator at high momentum is given by 
\begin{equation}
 \frac{1}{q^2} \ ,
\end{equation} 
while in the lattice formulation it reads
\begin{equation}
 \frac{1}{p^2} \ ,
\end{equation} 
where $p$ is the improved lattice momentum. 
However, the mapping from lattice to continuum allows to choose different definitions of the momentum, that differ from $q$ by 
higher order corrections in the lattice spacing. Recall that $p = q + \mathcal{O}(a^2)$. In general, the propagator is then given by
\begin{equation}
\frac{1}{\tilde{p}^{\, 2}}  \ ,
\end{equation}
with $\tilde{p} = q + \mathcal{O}(a^n)$ for $n \geq 2$. A possible choice for $\tilde{p}$ is then
\begin{eqnarray}
\tilde{p} & = &\frac{2}{a} \sin \frac{a \, q}{2}  \, + \, d_3 \, \left( \sin \frac{a \, q}{2} \right)^3 
\nonumber \\
& &   \, + \, d_5 \, \left( \sin \frac{a \, q}{2}  \right)^5 \, + \, d_7 \, \left( \sin \frac{a \, q}{2}  \right)^7 + \cdots
\nonumber \\
& = & p
+ \, p^3 \, \frac{a^2}{8} \, \left( a \,  d_3  - \frac{1}{3}\right)
\nonumber \\
& &
~  + \, p^5 \, \frac{a^4}{32} \, \left(-\frac{a}{2} \, d_3  +  a \, d_5 + \frac{1}{60}\right)
\nonumber \\
& &
~\quad
+ \, p^7 \, \frac{a^6}{128} \, \left(\frac{13 }{120} \, a \, d_3 - \frac{5}{6}  \, a \, d_5 + a \, d_7 - \frac{1}{2520}\right)
\nonumber \\
& &
~\quad\quad
+O\left(a^8\right)
\label{Eq.Expansao}
\end{eqnarray} 
where the coefficients $d_j$ can be adjusted to push the difference between $p$ and $\tilde{p}$ towards higher orders in $a$.
For example, the choice $d_3 = 1/ 3 \, a$, $d_5 = 3 / 20 \, a$, $d_7 = 5 / 56 \, a $ results in
$\tilde{p} = p + \mathcal{O}(a^7 \, q^8)$. From Eq. (\ref{Eq.Expansao}) it follows that the lattice propagator can be written as
\begin{equation}
\frac{1}{p^2} + \bar d_1 + \bar d_2 \, p^2 + \bar d_4 \, p^4 + \cdots
\end{equation}
The lattice propagator as written in Eq (\ref{Eq:EffParLat}) mimics this type of expression, with the leading term replaced by 
the one-loop continuum extrapolation lattice perturbation theory. The coefficients $c_i$ parametrize  in an effective way 
the finite size effects and, as in \cite{Ilgenfritz:2007qj}, they will be computed fitting the lattice data. 
Naively one expects that  the $c_i$'s decrease as the continuum limit is approached, i.e. as $\beta$ increases.

\begin{figure*}[t]
   \centering
   \includegraphics[scale=0.31]{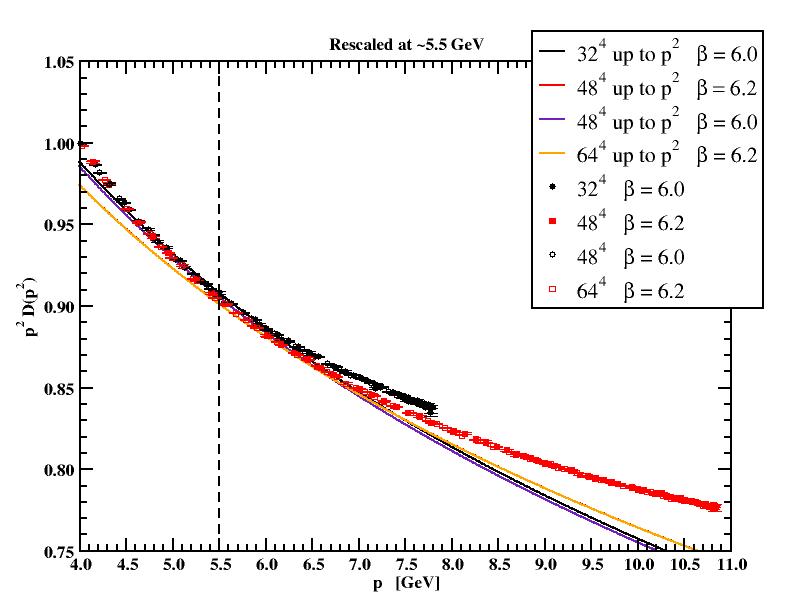}
   \includegraphics[scale=0.31]{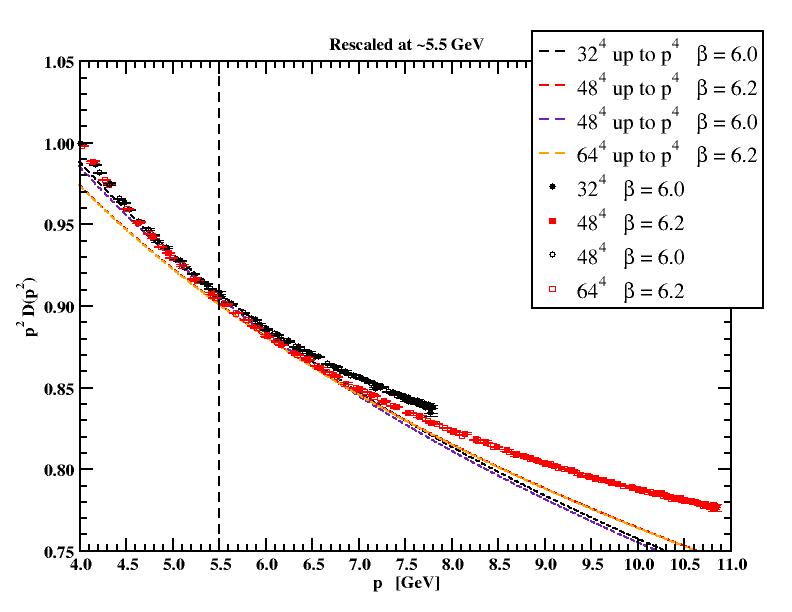}
   \caption{UV gluon dressing function corrected for the lattice artefacts. 
   The lines in the lhs figure refer to the outcome of the fits when  only terms up to $p^2$ are included to model the lattice artefacts. 
   The dashed lines in rhs figure were obtained including terms up to $p^4$ to model the lattice artefacts. See main text for details.}
   \label{fig:dresscorrected}
\end{figure*}

The outcome of fitting the lattice data for the largest possible momentum range with a $\chi^2/d.o.f. \sim 1$ is given in Tab. \ref{tab:EffArt}, when
the corrections to the one-loop term goes up to order $p^2$, see Eq. (\ref{Eq:EffParLat}), and in Tab. \ref{tab:EffArtp4}, when the corrections take into account terms
up to order $p^4$. It is only when the corrections of up to $p^4$ are taken into account that the range of momentum where the scale violation is observed,
i.e. for $p \gtrsim 5$ GeV, that Eq. (\ref{Eq:EffParLat}) is able to reproduce the lattice data over the full interval. For the intervals reported in
Tabs \ref{tab:EffArt} and \ref{tab:EffArtp4} the coefficients that effectively parametrize the lattice artefacts are small. There is a clear dependence on 
the lattice spacing, with the values reported for $Z$ and $c_j$'s being smaller for the smaller lattice spacing. Indeed, the $c_i$'s reported in 
Tab. \ref{tab:EffArtp4} are about one order of magnitude smaller for the results with the larger $\beta$ value. 
On the other hand, the quoted fitted values seem to be independent of the lattice volume.
Note that $c_1$ has dimensions GeV$^{-2}$, $c_2$ has dimensions of GeV$^{-4}$ and $c_3$ is given in GeV$^{-6}$.

One can assume that the corrected gluon dressing function is then given by the first term that appears in Eq. (\ref{Eq:EffParLat}), 
whose graphical representation can be seen in Fig. \ref{fig:dresscorrected} if one rescales $Z$ quoted in 
Tabs \ref{tab:EffArt} and \ref{tab:EffArtp4} to reproduce the lattice data at the extrema of the fitting interval.

The conclusion that can be drawn for this figure being that this way of estimating the lattice artefacts is able to provide estimations of the propagator
beyond the setting of the scale violation. Furthermore, the various estimations of the ``continuum propagator'' in the UV regime show a good
overlap, and the differences between the various curves provide an estimation of the theoretical uncertainty for the continuum propagator at high momentum.

\section{Summary and conclusions \label{Final}}

In this paper we discuss the scale setting problem in lattice simulations combined with the removal of the remaining finite size effects.

The lattice computation reported herein considers the Landau gauge gluon propagator at two different $\beta$ values and for large statistical ensembles of
gauge configurations. The ensembles considered in the current work have about 10$^4$ gauge configurations.
In a first  stage, the usual procedure to resolve the finite size effects, the introduction of momentum cuts, is considered. 
The momentum cuts are able to provide smooth curves for the lattice estimation of $D(p^2)$. 
The curves for the different  lattice setups are in good agreement, confirming that the momentum cuts are capable of handling 
the lattice effects efficiently. 

The procedure, i.e. the determination of a renormalized $D(p^2)$, requires a definition of the lattice spacing, which can introduce a
bias in the overall procedure. This allows to question if the agreement between the outcome of the two simulations can be improved by a different
choice of the lattice spacing. Indeed, by looking at the difference between the propagators and requiring it to be minimal over a range of momenta, it is possible to define
a relative lattice spacing between $\beta = 6.0$  and $\beta = 6.2$. Taking the coarser lattice as reference, it follows that the corresponding
value for $a$  for the fine lattice differs by less than 2\% from previous estimations. The devised calibration of the lattice spacing associated with different $\beta$'s
also helps in the reduction of possible bias introduced in the computation of $a$, that can come for the various sources that include also the use of gauge ensembles 
with quite different statistics. As observed, the differences in the values of $a$ from the various methods is below the $\sim$ 2 - 3\%. 
The overall agreement, within let us say 3\%, for the various estimations for the lattice spacing can also be viewed as an indication of the consistency of the lattice QCD approach to
non-perturbative physics.

Despite having improved the agreement between the different lattice simulations, the data shows, at high momentum, differences that come
from the finite size effects and, in principle, should be understood within perturbation theory. These scale violations in the UV regime have been observed
also in previous published lattice data. Taking the scaling violations has a measure of the finite size effects, then it is possible to claim a reliable propagator
for momenta up $\sim 5.5$ GeV. We call the reader attention, that the exact scale setting depends on the definition taken for the momentum 
and, certainly, the choice between using the naive momentum (\ref{p:naive}) or the improved momentum (\ref{p:improved}) impacts on the scale at 
which the scaling violations are observed.
Given that these violations occur in the UV regime, 
 the H(4) method is of limited help to handle such violations. 
 Indeed, the method requires, for each $p^2$, various $p^{[n]}$ to extrapolate the lattice data to the ``continuum'' limit and at large momenta, typically, there is a 
unique or few data points for each $p^2$, preventing the use of the method. In order to be able to use the H(4) method for high momenta, 
a functional form to model the corrections has to be introduced and a global fit to the data performed.
Herein, we explore the possibility of relying in perturbation theory, that is expected to apply in the regime where
the scale violations are seen, to resolve the finite size effects.

The perturbative inspired description of the high momenta region is always close to the renormalized $D(p^2)$. 
This is reassuring, as it is expected that perturbation theory holds at large momentum for an asymptotic free theory. 
However, perturbation theory is not able provide a satisfactory description of the lattice data in the region where the scale violations are observed. 
It is only when an effective description, that relies on one-loop inspired  results, is considered that the scaling violations seem to be resolved. 

The analysis of the lattice  data also rises the question of the comparison of continuum, understood as the perturbative description, and lattice results. 
Differences between the two types of approach have been observed, for example, for the plaquette. In our case, the reported differences become
clear due to the large statistical ensembles used, that exceed previous calculation by a factor of about 5. Our analysis find good agreement
between the results of the two approaches only after a proper handling of the finite size effects. Indeed, the reported results can be seen as
supporting the continuum predictions for momenta above $\sim 5$ GeV. From this point of view, the lattice data is compatible with one-loop
perturbative QCD results for momenta above this momentum scale.

A nice feature of the method described here to handle the finite size effects is that it can be generalized to resolve the lattice spacing effects in the UV regime to
other Green functions. In principle, the method has the potential to check for the consistency of the lattice approach and continuum perturbation theory beyond
two-point Green functions.

\begin{figure*}[t]
   \centering
   \includegraphics[scale=0.31]{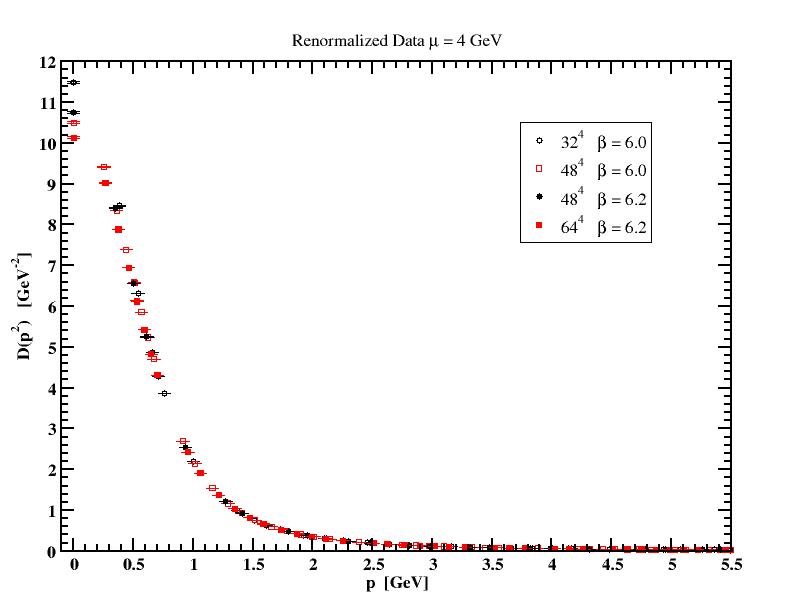} 
   \includegraphics[scale=0.31]{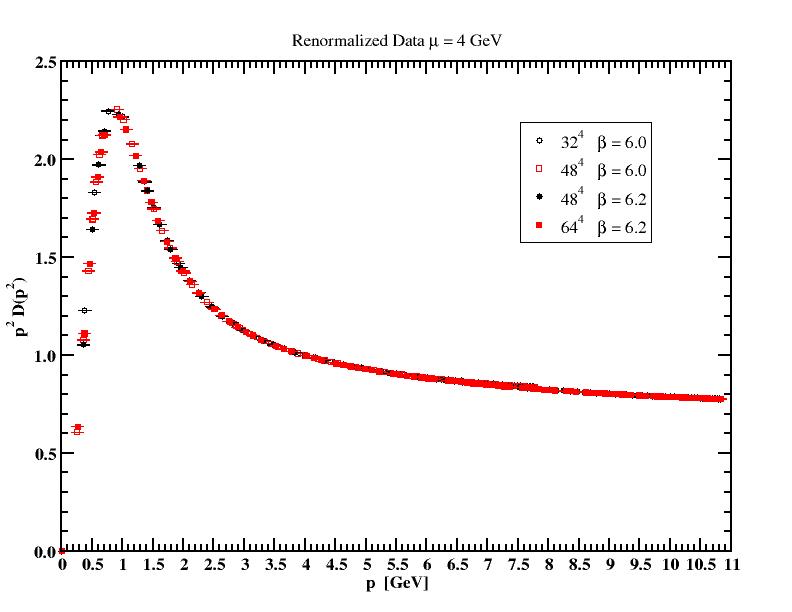}
   \caption{Gluon propagator up $p = 5.5$ GeV including all data below 700 MeV (left) and gluon dressing function for the full range of momenta.}
   \label{fig:glue}
\end{figure*}

To complement the information given so far, in Fig. \ref{fig:glue}  we show the outcome of the simulations for the gluon propagator and the gluon dressing function, 
using the relative setting of the lattice spacing, with the momentum cuts discussed for $p > 0.7$ GeV and including all the lattice data for smaller momenta  
as was done, for example,  in \cite{Dudal:2018cli}.
  Note 
   that the lattice data 
  shows finite volume effects that are observed in the deep infrared region. 
This outcome is inline with the conclusions reported in \cite{Oliveira:2012eh}.

\section*{Acknowledgements}

This work was financed through national funds by FCT - Fundação para a Ciência e Tecnologia, I.P. in the framework of the projects UIDB/04564/2020 , UIDP/04564/2020 and UID/04564/2025, with DOI identifiers  \url{10.54499/UIDB/04564/2020}, \url{10.54499/UIDP/04564/2020} and \url{10.54499/UID/04564/2025}.
P. J. Silva  acknowledges financial support from FCT contract CEECIND/00488/2017, with DOI identifier \url{10.54499/CEECIND/00488/2017/CP1460/CT0030}.
The authors acknowledge the Laboratory for Advanced
Computing at the University of Coimbra (\url{http://www.uc.pt/lca}) and the Minho Advanced
Computing Center (\url{http://macc.fccn.pt}) for providing access to the HPC resources. 
Access to Navigator was partly supported by the FCT Advanced Computing Projects 2021.09759.CPCA, 2022.15892.CPCA.A2 and 2023.10947.CPCA.A2 with DOI identifiers \url{10.54499/2021.09759.CPCA}, \url{10.54499/2022.15892.CPCA.A2} and  \url{10.54499/2023.10947.CPCA.A2} respectively. 
Access to Bob was supported by the FCT Advanced Computing Project CPCA/A2/6816/2020.
Access to Deucalion was supported by the
FCT Advanced Computing Project 2024.11063.CPCA.A3.



\end{document}